\documentstyle[aps,twocolumn,prb,epsf]{revtex} 
\begin{document}

\draft
\twocolumn[    

\hsize\textwidth\columnwidth\hsize\csname @twocolumnfalse\endcsname    

\title{Thermal transport in the Falicov-Kimball model on a Bethe lattice}

\author{A. V. Joura, D. O. Demchenko and J. K. Freericks}
\address{Department of Physics, Georgetown University, Washington, DC 20057-0995, U.S.A.}

\date{\today}
\maketitle

\widetext
\begin{abstract}

We calculate thermal transport in the Falicov-Kimball model on an infinite-coordination-number Bethe lattice. 
We perform numerical calculations of the thermoelectric 
characteristics and
concentrate on finding materials parameters for 
which the electronic thermoelectric figure-of-merit $ZT$ is large, suggesting potential 
cooling and power generation applications. 
Surprisingly, the Bethe lattice has significant qualitative and quantitative differences with 
the previously studied hypercubic lattice. 
At low temperature 
it is unlikely that these systems can be employed in thermoelectric devices
due to the low conductivities and due to a larger lattice contribution to the thermal conductivity $\kappa_L$, but at 
high temperature, the thermoelectric parameters appear more promising for devices
due to a significant enhancement of $ZT$ and a smaller relative contribution by the 
lattice thermal conductivity. 

\end{abstract}
\pacs{PACS numbers: 72.15.Jf; 72.20.Pa, 71.27.+a, 71.10.Fd}
]      

\narrowtext

Materials that are candidates for thermoelectric applications 
\cite{DiSalvo}
are interesting because they
have the potential to compete with conventional compressor-based coolers. These materials 
are characterized by the thermoelectric figure-of-merit $ZT$, which 
evaluates the efficiency of a bulk material as a cooling element. Commercially available 
semiconductor devices exhibit a $ZT$ of about unity. Such values of $ZT$ are 
far too low 
to be competitive with the efficiency of conventional
mechanical refrigerators, which would require a $ZT$ of 3 or 4.


Current directions of research include bulk semiconductor devices \cite{semi_bulk,low_T}, 
semiconductor nanostructures \cite{semi_nano}, 
metal/correlated semiconductor heterostructures \cite{Sham},
and strongly correlated materials \cite{Mahan,freericks_zlatic,hc_our_prb,Kondo}. 
Bulk semiconductor devices are most commonly employed, and $ZT$s of up to 1.14 
have been achieved for a $p$-type
(Bi$_2$Te$_3)_{0.25}$(Sb$_2$Te$_3)_{0.72}$(Sb$_2$Se$_3)_{0.03}$ alloy at 
room temperature and pressure \cite{semi_bulk}. Below room temperature, 
however, a maximum $ZT$ of only 0.8 has been found in bulk \cite{low_T}
CsBi$_4$Te$_6$ at 225 K.
Semiconductor nanostructures appear more promising 
by increasing 
$ZT$ due to a suppressed lattice contribution to 
the thermal conductivity. The value of 2.4 has been observed experimentally 
\cite{semi_nano} for $p$-type Bi$_2$Te$_3$/Sb$_2$Te$_3$ superlattice 
structures at 300 K. 
These structures also showed improved values of $ZT$ ($\sim 1.7$) at 
temperatures as low as 210 K. 
Nevertheless, at lower temperatures all measured bulk materials exhibit $ZT$s 
that are lower than unity. A recent theoretical proposal predicts an enhanced $ZT$
in a metal/correlated semiconductor structure \cite{Sham}. 
Estimated values of the low temperature $ZT$s are enormous; for instance, for 
a semiconductor gap of 100 meV, the best value of $ZT$
would be $\sim$ 6 at 150 K and $\sim$ 100 at 40 K. 
Finally, strongly correlated materials have been suggested as possible
candidates for bulk thermoelectric materials, especially at low temperatures
\cite{Mahan,freericks_zlatic,hc_our_prb,Kondo} because the correlations can strongly 
renormalize the Fermi temperature to low temperatures. 
Very recently, Peltier 
cooling has been achieved experimentally \cite{Kondo} below 10 K using 
crystals of the Kondo metal CeB$_6$. 
The greatest value of cooling at $T = 4.5$ K was equal to 0.2 K, 
corresponding to a $ZT$ of about 0.26. 
However, the properties of such correlated materials are still not well characterized, 
namely, whether large values of $ZT$ predicted theoretically can be 
realized experimentally or if they are the result of unavoidable simplifications.

In this paper, we discuss electronic (thermal and charge) transport
for the Falicov-Kimball model on an infinite-coordination-number Bethe lattice. 
In a recent article \cite{hc_our_prb}, we investigated transport on an infinite-dimensional 
hypercubic lattice 
and found possible sets of parameters where $ZT > 1$ for a wide range of temperatures.
However, some of the Bethe lattice's properties 
are closer to those of real three-dimensional systems.
First of all, the electronic band structure for the Bethe lattice has a finite bandwidth, 
as opposed to the hypercubic
lattice where the interacting density of states (DOS) decreases
exponentially.
Another important distinction is
that the ``quasiparticle'' scattering time 
$\tau(\omega)$ for the Bethe lattice is exactly zero
whenever the interacting DOS is zero. This is also expected for a real 
physical system, in contrast to the hypercubic lattice \cite{hc_our_prb} where the
scattering time
approaches a non-zero constant as $\omega\rightarrow\pm\infty$ and behaves as a 
power law at small $\omega$ in the ``pseudo gap''. 
The Bethe lattice, however, also has some unphysical properties.
For instance, the description of transport is controversial \cite{Blumer} for the Bethe lattice; 
here we choose to define the square of the velocity of a
``quasiparticle'' in such a way that the optical sum rule is enforced \cite{sum_rule}. 
There also is no apparent way to determine the volume for the Bethe lattice, 
which leads to the impossibility of a microscopic determination of the electric conductivity unit.

We consider the Hamiltonian for the spinless Falicov-Kimball 
model \cite{falicov_kimball}
\begin{equation}
H=-\frac{t^*}{\sqrt{Z}}\sum_{\langle i,j\rangle}c^\dagger_{i}
c_{j}+U\sum_{i}c^{\dagger}_{i}c_{i}f^{\dagger}_{i}f_{i},
\label{eq: ham}
\end{equation}
where $c^{\dagger}_{i}$ ($c_{i})$ and $f^{\dagger}_{i}$ ($f_{i})$ are the 
conduction and localized electron creation 
(annihilation)
operators respectively for a spinless electron at site $i$, 
and $U$ is the on-site Coulomb
interaction strength.  The hopping integral is scaled with the number of nearest
neighbors $Z$ so as to have a finite result in the limit \cite{metzner_vollhardt,Kotliar} of
$Z\rightarrow \infty$; 
we measure all energies in units of $t^*=1$. 
As shown earlier, the case of infinite $Z$ provides 
an opportunity 
to take advantage of the locality of the self-energy and to use dynamical
mean field theory (DMFT) (for a review see Refs.\onlinecite{Kotliar,freericks_review}).
Here we consider the Bethe lattice so the noninteracting density of states is
a semicircle 
$\rho(\epsilon)=\sqrt{4-\epsilon^2}/2\pi$. 
The method for solving the Falicov-Kimball model using DMFT is given 
elsewhere\cite{freericks_review,brandt_mielsch,freericks_fk}. 
We work with a fixed value $w_1=\langle n_f\rangle$ for the average number of localized electrons,
which is the so-called binary alloy picture. 
The other two input parameters needed to calculate the Green's 
function $G(\omega)$ and self energy $\Sigma(\omega)$ 
are $U$ and $\rho_e$ $-$ the average number of conduction electrons. 
All our calculations are performed for the case $\rho_e=1-w_1$. 
This case is of particular interest because it yields a correlated
insulating state at values of $U$ larger than a certain critical $U_c$.


The local Green's function $G = G(\omega)$ on the real axis
satisfies the following cubic 
equation\cite{vanDongen_92,vandongen_leinung}:

\begin{eqnarray}
G^3-2(\omega+ \mu-U/2)G^2+[1+(\omega+\mu-U/2)^2
\cr
-{U^2\over 4}]G-[\omega+ \mu-U/2+U(w_1-{1\over2})]=0
\label{eq: cubic1}
\end{eqnarray}
which we solve numerically in order to determine the Green's function, and 
subsequently, the self-energy from the relation (valid on the Bethe lattice):
\begin{equation}
\Sigma(\omega)=\omega+\mu-G(\omega)-{1\over G(\omega)}.
\label{eq: self-energy}
\end{equation}
Note that the physical root must be chosen in Eq. (\ref{eq: cubic1}) to yield the retarded Green's function
and a chemical potential $\mu$ is employed to get the right average electron density $\rho_e$. 

In order to find the transport,
we start from the exact expression for the scattering time 
on the Bethe lattice \cite{sum_rule}
\begin{equation}
\tau(\omega)=\frac{1}{3} \int d\epsilon \rho(\epsilon)(4-\epsilon^2)A^2(\epsilon,\omega), 
\label{tau_definition}
\end{equation}
where $A(\epsilon,\omega)=-\frac{1}{\pi}\textrm{Im}\frac{1}{\omega+\mu-\Sigma(\omega)-\epsilon}$ 
is the spectral function. 
This expression can be 
evaluated in terms of the local Green's function as
\begin{equation}
\tau(\omega)={1\over 3\pi^2}\textrm{Im}^2[G(\omega)]{|G(\omega)|^2-3\over|G(\omega)|^2-1}.
\label{tau}
\end{equation}
The scattering time computed on the Bethe lattice possesses several important properties.
It resembles the interacting density of states in shape and behavior as a function of the Coulomb 
interaction $U$ and filling $w_1$. Namely, it develops a well-defined gap at the metal-insulator
transition, as opposed to the hypercubic lattice case, where $\tau(\omega)$ 
assumes a power law behavior.
In addition, 
on the Bethe lattice, the relaxation time is equal to 
zero outside of the band for large $|\omega|$, whereas $\tau(\omega)$ on the 
hypercubic lattice approaches a nonzero constant. There is, however, an
essential difference, namely, it can be shown that in the vicinity of the band 
edges at zero temperature $\tau(\omega)$ obeys
\begin{eqnarray}
\tau(\omega)\approx A_{\tau}\theta(\omega-{E_g\over2})(\omega-{E_g\over2}) 
\cr
+ B_{\tau} \theta(-\omega-{E_g\over2})(-\omega-{E_g\over2}),
\label{tau_edge}
\end{eqnarray}
where $A_{\tau}$ and $B_{\tau}$ are constants, $\theta(x)$ is the unit step-function
and $E_g$ is the bandgap. 
On the other hand, the interacting DOS 
[$\rho_{int} (\omega)\equiv -\frac{1}{\pi}\textrm{Im}G(\omega)$]
at $T=0$ behaves as \cite{vandongen_leinung}
\begin{eqnarray}
\rho_{int}(\omega)\approx A\theta(\omega-{E_g\over2})\sqrt{|\omega-{E_g\over2}|}
\cr
+ B \theta(-\omega-{E_g\over2})\sqrt{|-\omega-{E_g\over2}|},
\label{dos_edge}
\end{eqnarray}
where $A$ and $B$ are again constants. 
Note, that at finite temperatures
the expressions in Eqs. (\ref{tau_edge}) and (\ref{dos_edge}) are modified 
by a temperature-dependent shift in the chemical potential ($\omega \rightarrow \omega+\mu - \mu_{T=0}$).
This difference in behavior of $\tau(\omega)$ and $\rho_{int}(\omega)$
in Eqs. (\ref{tau_edge}) and (\ref{dos_edge}) arises from the extra velocity 
squared terms in $\tau(\omega)$ which modify the behavior at the band edges. 
Hence, $\tau(\omega)\ne \tau_0\rho_{int}(\omega)$ which would be the simplest approximation.

Once $\tau(\omega)$ is known, we can compute the transport coefficients $L_{ij}$, 
according to the Jonson-Mahan theorem \cite{jonson_mahan} 
(see also Ref. \onlinecite{freericks_zlatic}):
\begin{equation}
L_{ij}=\frac{\sigma_0}{e^2}\int_{-\infty}^{\infty}d\omega
\left ( -\frac{df(\omega)}{d\omega} \right ) \tau(\omega)\omega^{i+j-2},
\label{eq: lij_final}
\end{equation}
where $\sigma_0$ has units of conductivity $e^2/ha$, and $a$ 
is a length scale that
cannot be independently determined on the Bethe lattice. 
Here, $f(\omega)=1/[1+\exp(\omega/k_BT)]$ is the Fermi-Dirac 
distribution.
Thermoelectric 
characteristics can be obtained once these transport coefficients are
determined. 
Thus,
\begin{equation}
\sigma_{dc}=e^2L_{11},
\label{eq: conductivity}
\end{equation}
\begin{equation}
S=-\frac{k_B}{|e|T}\frac{L_{12}}{L_{11}},
\label{eq: thermopower}
\end{equation}
\begin{equation}
\kappa_e=\frac{k_B^2}{T}\left [ L_{22}-\frac{L_{12}L_{21}}{L_{11}}\right ],
\label{eq: thermalconductivity}
\end{equation}

\begin{equation}
ZT=\frac{L_{12}^2}{L_{11}L_{22}-L_{12}^2},
\label{eq: fom_final}
\end{equation}
and
\begin{equation}
{\mathcal L}
=\left (\frac{e}{k_B}\right )^2\frac{\kappa_e}{\sigma_{dc}T}=
\frac{L_{11}L_{22}-L_{12}^2}{L_{11}^2T^2}.
\label{eq: lorenz_final}
\end{equation}
Here, $\sigma_{dc}$ and $\kappa_e$ are the electric and thermal (electronic part) 
conductivities respectively,
$S$ is the thermopower, $ZT$ is the thermoelectric figure-of-merit, and ${\mathcal L}$ is
the Lorenz number. Note that the lattice contribution to the thermal conductivity has 
been neglected here.

At low temperature, in the correlated insulator, we can determine analytic 
expressions for the electronic transport.
First, note that $\Delta\mu=\mu - \mu_{T=0}$, the change in chemical 
potential with temperature, is a linear function to leading order 
\begin{equation}
\Delta\mu\approx {T\over2}\textrm{ln}({B\over A})+O(T^{3\over2}).
\label{eq: delta_mu}
\end{equation}
Let us denote 
\begin{equation}
a=A_\tau\sqrt{B\over A} \ \ \ 
b=B_\tau\sqrt{A\over B},
\label{eq: denote}
\end{equation}
then the thermoelectric characteristics follow as
\begin{equation}
\sigma_{dc}\approx\sigma_0e^{-{\beta E_g\over2}}[T(a+b)+O(T^{3\over2})],
\label{eq: sigma_app}
\end{equation}
\begin{equation}
S\approx-{k_B\over |e|}\cdot{a-b\over a+b}\cdot{E_g\over 2T}+O(T^{-{1\over2}}),
\label{eq: S_app}
\end{equation}
\begin{equation}
\kappa_e\approx {\sigma_0k_B^2\over e^2}e^{-{\beta E_g\over2}}
[{ab\over a+b} E_g^2 + O(T^{1\over2})],
\label{eq: ke_app}
\end{equation}
\begin{equation}
{\mathcal L}\approx{\sigma_0\over e^2}{ab\over (a+b)^2}\Bigl({E_g\over T}\Bigr)^2 + O(T^{-{3\over2}}),
\label{eq: L_app}
\end{equation}
and
\begin{equation}
ZT\approx{(a-b)^2\over 4ab} + O(T^{1\over2}).
\label{eq: ZT_app}
\end{equation}
Therefore, at low temperatures, the thermopower $S$ and Lorenz number $\mathcal L$ diverge
as $1/T$ and $1/T^2$, respectively.  Then, $\sigma_{dc}$ and $\kappa_e$
assume exponentially small values, and $ZT$ approaches a nonzero constant 
(in the metallic phase $ZT$ vanishes as $T^2$). 

A divergence of the thermopower does not necessarily lead to unphysical consequences. 
For instance, 
the Peltier heat between two materials satisfies $Q_p=-T(S_2-S_1)j$, with $j$ 
the electrical current flowing from material 1 to material 2.
Even if the Peltier 
coefficient $-T(S_2-S_1)$ is finite as $T \rightarrow 0$ the Peltier heat vanishes because the current
decreases exponentially with $T$ as $T \rightarrow 0$.

\begin{figure}
\epsfxsize=3.0in
\centerline{\epsffile{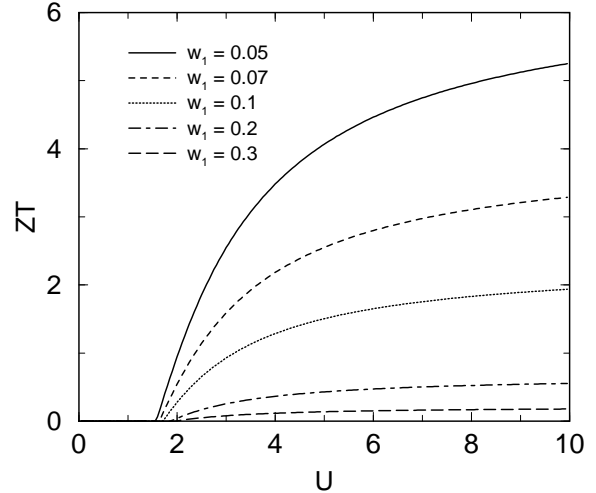}}
\caption{Thermoelectric figure-of-merit as a function of $U$ at $T = 0$ for
the Falicov-Kimball model at $\rho_e=1-w_1$, and $w_1=0.05$ (solid), 
0.07 (dashed), 0.1 (dotted), 0.2 (chain-dotted), and 0.3 (long-dashed).
\label{fig: ZT_U}}
\end{figure}

\begin{figure}[htbf]
\epsfxsize=3.0in
\centerline{\epsffile{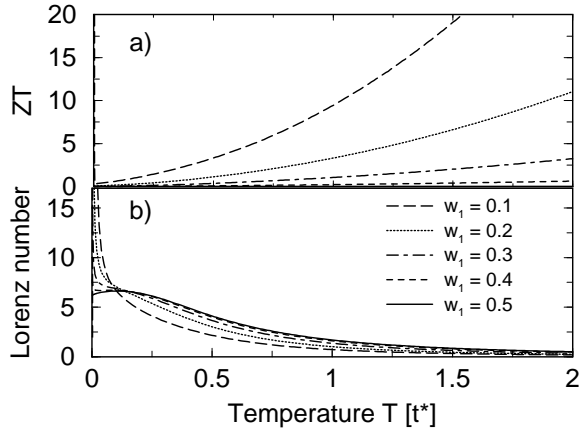}}
\caption{(a) Thermoelectric figure-of-merit and (b) the Lorenz number for
the Falicov-Kimball model at $U=2$, $\rho_e=1-w_1$, and $w_1=0.5$ (solid), 
0.4 (dashed), 0.3 (chain-dotted), 0.2 (dotted), and 0.1 (long-dashed).
\label{fig: ZT_L}}
\end{figure}

Now we show our numerical results.
In Fig. \ref{fig: ZT_U} we present the results for the thermoelectric figure-of-merit $ZT$ at $T=0$ as a 
function of $U$ for several different values of $w_1$. 
For small $U$ (in a metal) 
$ZT$ vanishes as $T \rightarrow 0$. As the gap opens, $ZT$ starts to grow linearly with $U$. It is clear
that the asymmetry of the interacting DOS, which increases as $w_1$ differs from 0.5
brings about larger values 
of $ZT$ for a given $U$. The closer $w_1$ is to zero or unity the larger $ZT$ is. 
Our numerical results suggest that zero temperature values of $ZT$ can be made 
as large as desired by choosing proper values of $U$ and $w_1$. This result is, however,
undermined by the fact that conductivity is exponentially small for low 
temperatures [Eq. (\ref{eq: sigma_app})], so high voltage would be needed to operate a device. 
Note that on the hypercubic lattice \cite{hc_our_prb} $ZT \rightarrow 0$ as $T \rightarrow 0$
in both the metal and the insulator. 
Figure \ref{fig: ZT_L} shows the thermoelectric figure-of-merit and the Lorenz number 
as functions of temperature for $U=2$. At half-filling $U=2$ is the critical value 
of the Coulomb interaction for the metal-insulator transition and the critical $U_c$ decreases
as $w_1$ changes from 0.5. (For other values of $U>U_c$ the 
temperature dependence of all thermoelectric parameters behaves similarly, with
only quantitative differences.) The thermoelectric figure-of-merit grows monotonically 
with temperature and reaches large values at high temperature. 
The low temperature peak found close to half-filling on the hypercubic lattice \cite{hc_our_prb}
does not occur on the Bethe lattice.
The largest values of $\mathcal L$ are achieved at low temperatures in the insulating phase,
while in the metallic phase, the Lorenz number has a maximum at finite temperatures.
\begin{figure}[htbf]
\epsfxsize=3.0in
\centerline{\epsffile{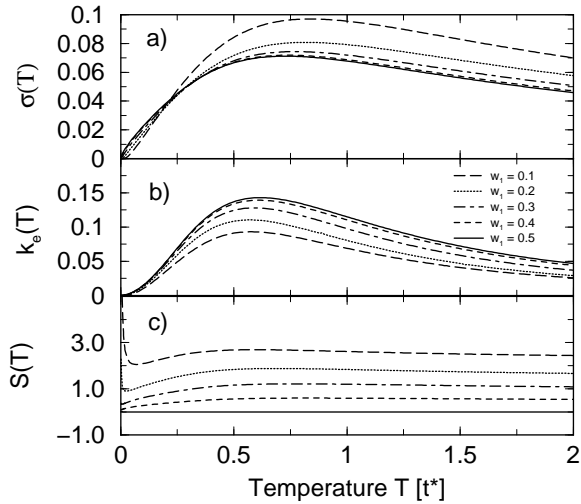}}
\caption{(a) Electric conductivity, (b) electronic contribution to the thermal 
conductivity, and (c) thermopower for the
Falicov-Kimball model at $U=2$, $\rho_e=1-w_1$, and $w_1=0.5$ (solid), 
0.4 (dashed), 0.3 (chain-dotted), 0.2 (dotted), and 0.1 (long-dashed).
\label{fig: Sigma_ke_S}}
\end{figure}

The temperature dependence of the electric conductivity $\sigma_{dc}$, the electronic part
of the thermal conductivity $\kappa_e$, and the thermopower $S$ for $U=2$ and various $w_1$
are shown in Fig. \ref{fig: Sigma_ke_S}. 
Both $\sigma_{dc}$ and $\kappa_e$ are exponentially small at low temperatures in agreement 
with the analytical expressions in Eqs. (\ref{eq: sigma_app}) and (\ref{eq: ke_app}), reach
a maximum, and then decrease at higher temperatures. Larger maximum values 
of $\kappa_e$ are achieved closer to half-filling, contrary to the behavior of $\sigma_{dc}$, where
larger maximum values are reached away from $w_1=0.5$. The thermopower increases away from 
half-filling as well, and exactly vanishes for $w_1=0.5$ (due to particle-hole symmetry). 
It has a pole at $T=0$, according to Eq. (\ref{eq: S_app}).
Also, the thermopower is positive,
indicating hole-like transport for $w_1<0.5$, and it satisfies the relation 
$S(w_1,U,T)=-S(1-w_1,U,T)$ from particle-hole symmetry.  

A comment is in order as to the influence of the neglected lattice component 
of the thermal conductivity. 
At low temperatures ($T$ well below the Debye temperature $\Theta_{D}$) 
the lattice contribution to the thermal conductivity $\kappa_L$
decreases \cite{ashcroft} as $T^3$. 
At the same time, the electronic contribution $\kappa_e$ is exponentially small, 
according to Eq. (\ref{eq: ke_app}). As a consequence, the lattice
component is likely to dominate the thermal conductivity and, therefore,
will reduce the value of $ZT$. On the other hand, at high temperatures 
($T \gg \Theta_{D}$) the rate of decline of the lattice component 
is \cite{ashcroft} $\kappa_L \sim {1/T^x}$, where
$x$ is between 1 and 2. The derivation of the precise behavior is rather 
complex, related to the competition between cubic and quartic 
anharmonic scattering processes \cite{Herring}.
The important point, however, is that at the same time $\kappa_e$ 
assumes large values and should become dominant, 
allowing the high values of $ZT$ to be experimentally achievable.

In summary, we have analyzed the thermal transport properties of strongly correlated 
materials within the Falicov-Kimball model on the Bethe lattice.
Significant values of $ZT$ can be obtained for both low
and high temperature regimes, contrary to the hypercubic lattice case, 
where $ZT$ declines at high temperatures. On the other hand, 
$\kappa_e$ grows linearly at high temperature on the hypercubic lattice, while
on the Bethe lattice it decreases. 
The behavior of the interacting density of states $\rho_{int}(\omega)$ 
as well as that of the scattering time $\tau(\omega)$
suggest that the results obtained on the Bethe lattice are closer to what happens 
in real materials. 
However, while high $ZT$ values are necessary for 
practical applications, they are not sufficient to guarantee a viable device. 
In order to take advantage of these
properties, one needs to be able to achieve appreciable values of the current 
densities. Since the conductivity approaches zero for $T\rightarrow 0$, 
the feasibility of thermoelectric devices based on strongly correlated materials
described by this Falicov-Kimball model is questionable at low temperatures. 
At high temperatures, judging by the high values of $ZT$ and lower lattice 
contribution to the thermal conductivity, 
it may be possible to employ these materials for power generation applications.

\acknowledgments
We would like to acknowledge the support of the National Science Foundation, 
grants DMR-9973225 and DMR-0210717, and the Office of Naval Research, grant 
number N00014-99-1-0328. 


\end{document}